\documentclass[12pt,twocolumn,tightenlines,aps,amsmath,floatfix,notitlepage,nofootinbib,prb]{revtex4}
\usepackage[T1]{fontenc}        
\usepackage{palatino}           
\usepackage{amsmath}            
\usepackage{amssymb}            
\usepackage{graphicx}           
\usepackage{rotating}
\usepackage{booktabs}
\usepackage{multirow}
\usepackage{pslatex}
\usepackage{epsfig}
\usepackage{epsf}
\usepackage[usenames]{color}
\usepackage{mathrsfs}           
\newcommand{\sfrac}[2]{\frac{#1}{#2}}
\usepackage{caption}
\newcommand{\V}[1]{\mathbf{#1}}
\newcommand{\bra}[1]{\langle #1|}
\newcommand{\ket}[1]{|#1\rangle}

\newcommand{\rem}[1]{}          
\newcommand{\mc}[3]{\multicolumn{#1}{#2}{#3}}

\begin{document}

\title{Magnetic and electrical properties of (Pu,Lu)Pd$_3$}

\author{M. D. Le} 
\affiliation{Helmholtz-Zentrum Berlin f\"ur Materialen und Energie, Hahn-Meitner-Platz 1, D-14109 Berlin, Germany}
\email{duc.le@helmholtz-berlin.de}
\author{K. A. McEwen}
\affiliation{Department of Physics and Astronomy, and London Centre for Nanotechnology, University College London, WC1E 6BT, UK}
\author{E. Colineau, J.-C. Griveau, and R. Eloirdi}
\affiliation{European Commission, Joint Research Centre, Institute for Transuranium Elements, Postfach 2340, 76125 Karlsruhe, Germany}

\date{\today}

\begin{abstract}

We present measurements of the magnetic susceptibility, heat capacity and electrical resistivity of Pu$_{1-x}$Lu$_x$Pd$_3$, with $x$=0, 0.1, 0.2, 0.5,
0.8 and 1. PuPd$_3$ is an antiferromagnetic heavy fermion compound with $T_N=24$~K. With increasing Lu doping, both the Kondo and RKKY
interaction strengths fall, as judged by the Sommerfeld coefficient $\gamma$ and N\'eel temperature $T_N$. Fits to a crystal field model of the
resistivity also support these conclusions. The paramagnetic effective moment $\mu_{\mathrm{eff}}$ increases with Lu dilution, indicating a decrease
in the Kondo screening. In the highly dilute limit, $\mu_{\mathrm{eff}}$ approaches the value predicted by intermediate coupling calculations. In conjunction with
an observed Schottky peak at $\sim$60~K in the magnetic heat capacity, corresponding to a crystal field splitting of $\sim$12~meV, a mean-field
intermediate coupling model with nearest neighbour interactions has been developed.

\end{abstract}


\maketitle
%
%

\section{Introduction}

The AnPd$_3$ series of compounds, with An=U, Np, or Pu, are rare examples of actinide intermetallic compounds in which the 5$f$ electrons are well
localised around the ionic sites. UPd$_3$ is a very interesting compound which exhibits four quadrupolar ordered phases below 8K~\cite{hcwESRF07},
whilst there are indications that NpPd$_3$ may also show quadrupolar order at low temperatures~\cite{hcw07}. These two compounds crystallise in the
double-hexagonal close-packed (dhcp) structure, in contrast to PuPd$_3$ which adopts the AuCu$_3$ structure, with lattice parameter $a=$4.105\AA. This
reflects the increasing localisation of the $5f$ electrons as shown by recent photoelectron spectroscopy measurements~\cite{Gouder06}.

Early measurements of the bulk properties and neutron diffraction studies~\cite{NHL+74} show it to be antiferromagnetic, with a transition temperature
$\sim$24~K, and a G-type structure, where nearest neighbour moments are aligned antiparallel. The same study found that the high temperature
resistivity shows a Kondo-like behaviour, increasing with decreasing temperature. This behaviour, together with a high Sommerfeld coefficient deduced
from recent heat capacity measurements~\cite{le07pupd3}, led us to make a further investigation of the properties of PuPd$_3$. 

Our aim has been to study how the competition between the Kondo effect and the RKKY exchange interaction affects the physical properties of this
compound. This may be accomplished by doping with a non-magnetic ion, which increases the distance between localised $f$-electrons, and hence
decreases the RKKY interaction. Finally, the single-ion properties of the Pu ion may be investigated in the highly dilute limit.

In this work, we present in section~\ref{sec-pupd3-exp} the experimental details and in section~\ref{sec-magcp} the measurements of the 
magnetic susceptibility, and heat capacity of
Pu$_{1-x}$Lu$_x$Pd$_3$. These measurements are then analysed using a localised moment mean field model in section~\ref{sec-numcalcs}. Finally,
section~\ref{sec-elec} presents measurements of the electrical resistivity and Hall coefficient of Pu$_{1-x}$Lu$_x$Pd$_3$, which are assessed in terms
of a simple crystal field model.

\section{Experimental Details} \label{sec-pupd3-exp}

\begin{figure}
  \begin{center}
    \includegraphics[width=1.0\columnwidth]{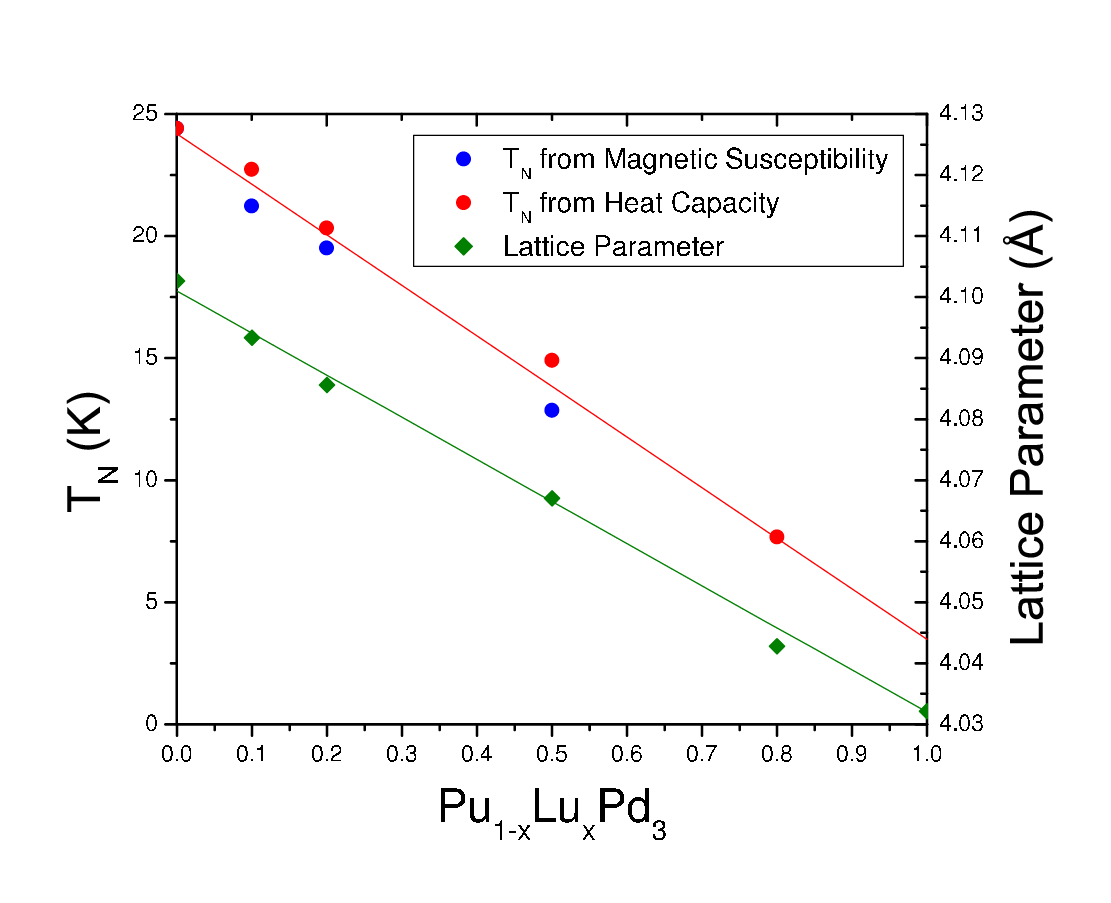}
    \caption{(Color online) Lattice parameters and transition temperatures for Pu$_{1-x}$Lu$_x$Pd$_3$.} \label{fg:pulupd3latticepar}
  \end{center}
\end{figure}

Polycrystalline samples of PuPd$_3$, LuPd$_3$ and Pu$_{1-x}$Lu$_x$Pd$_3$, with $x=0.1$, 0.2, 0.5, and 0.8, were produced at ITU by arc melting
appropriate amounts of the constituent elements under a high purity argon atmosphere on a water-cooled copper hearth, using a Zr getter. The AuCu$_3$
structure was confirmed by x-ray diffraction for each sample, and the lattice parameters are shown in figure~\ref{fg:pulupd3latticepar}. The data
show a linear dependence of the lattice parameter with increasing Lu dilution, in accordance with Vegard's Law, which also confirms the stoichiometry
of the samples. In addition, there also appears to be a linear dependence of the transition temperature, $T_N$, with doping. These temperatures were
determined from magnetic susceptibility and heat capacity measurements described in the next section. At $T_N$ there is a maximum in the susceptibility,
$\chi$, and hence this temperature was determined by numerically differentiating the data to find $\frac{d\chi}{dT}=0$. In the heat capacity, $C_p$, there is
a lambda step at $T_N$, which was determined by differentiating the data to find the minima of $\frac{dC_p}{dT}$. The values of $T_N$ deduced from
these two measurements are in close agreement, whereas the inflexion points of the resistivity data do not correlate with $T_N$ as determined from
$\chi$ or $C_p$. Nevertheless, the resistivity inflexion points show the same decreasing trend with Lu-doping as $T_N$. Errors in $T_N$ quoted in
table~\ref{tab:pulupd3cpcwinfo} were determined by the width in temperature of the lambda step for $C_p$ or the step in $\frac{d\chi}{dT}$.

The magnetisation and susceptibility were measured using a SQUID magnetometer (Quantum Design MPMS-7), whilst the heat capacity was determined by the
hybrid adiabatic relaxation method in a Quantum Design PPMS-9 for PuPd$_3$ and LuPd$_3$, and in a PPMS-14 for Pu$_{1-x}$Lu$_x$Pd$_3$. Small
samples with mass less than 5~mg were used for the heat capacity measurements so that the decay heat does not significantly affect the measurements. 
The X-ray diffraction, magnetisation and heat capacity measurements were made immediately after the preparation of samples in order to minimise the 
effects of radiation damage.

Finally, thin parallel-sided samples of each composition were extracted, polished and mounted for electrical transport measurements. As these
measurements were made some three months after synthesis, we observed significant radiation damage which manifested in a high residual resistivity at
low temperatures. This prompted us to anneal the samples at 800$^{\mathrm{o}}$~C for 12~h, and to remeasure the electrical transport properties. The
remeasured data is presented in section~\ref{sec-elec}.

\section{Magnetisation and Heat Capacity Measurements} \label{sec-magcp}
\subsection{Magnetisation} \label{sec-pupd3-expMag}

\begin{figure}
  \begin{center}
    \includegraphics[width=0.9\columnwidth]{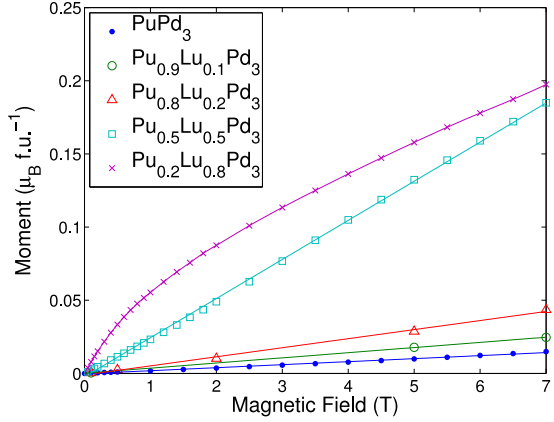}
    \caption{(Color online) The magnetisation at 2~K of Pu$_{1-x}$Lu$_x$Pd$_3$. Lines are guide to the eye.} \label{fg:pulupd3magnetisation} 
  \end{center}
\end{figure}

\begin{figure}
  \begin{center}
    \includegraphics[width=0.9\columnwidth]{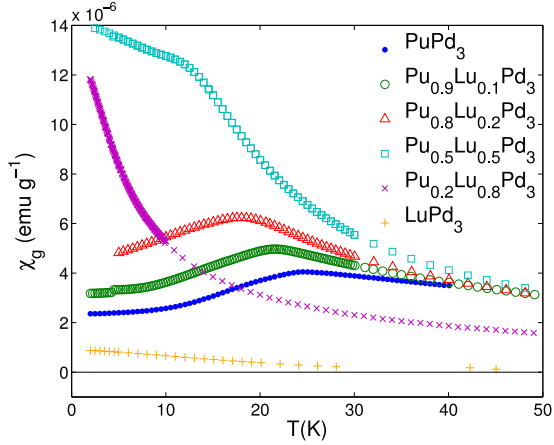}
    \caption{(Color online) The magnetic susceptibility of Pu$_{1-x}$Lu$_x$Pd$_3$.} \label{fg:pulupd3susc} 
  \end{center}
\end{figure}

\begin{figure}
  \begin{center}
    \includegraphics[width=0.91\columnwidth]{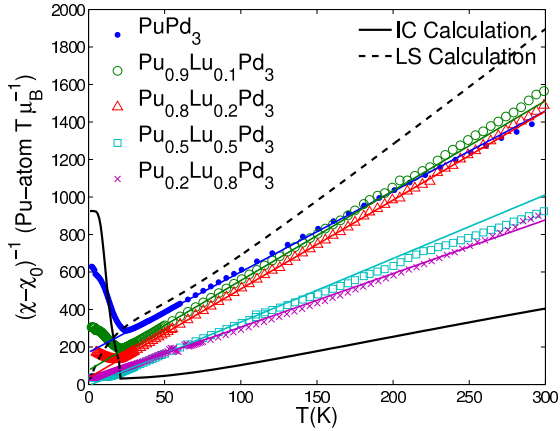}
    \caption{(Color online) The inverse magnetic susceptibility of Pu$_{1-x}$Lu$_x$Pd$_3$. Thin solid lines are fits to the modified Curie-Weiss Law for each compound. 
     The mean field intermediate coupling calculation is shown as a thick solid line, whilst the dashed line shows the single ion susceptibility calculated 
     from the crystal field in $LS$-coupling.}
     \label{fg:pulupd3invchi}
  \end{center}
\end{figure}

Figure~\ref{fg:pulupd3magnetisation} shows the magnetisation at 2~K, which shows that the Pu-rich compositions are not saturated at 7~T. This is not
surprising because we expect a $J=5/2$ ion to be saturated at a field $\gtrsim 100$~T, when the splitting between the lowest two CF levels is
$\gg$2~K. The magnetisation of Pu$_{0.2}$Lu$_{0.8}$Pd$_3$ shows some evidence of saturation, however.
The magnetic susceptibility is shown in figure~\ref{fg:pulupd3susc}, and the inverse susceptibility in figure~\ref{fg:pulupd3invchi}. The data in the 
paramagnetic phase above the N\'eel temperature are well fitted by a modified Curie-Weiss Law

\begin{equation} \label{eq:curieweiss}
\frac{M}{H} = \frac{N \mu_{\mathrm{eff}}^2\mu_B^2}{3k_B}\frac{1}{T-\theta_{\rm CW}} + \chi_0(H)
\end{equation}

\noindent where $N$ is the number of Pu atoms in the compound, and $\theta_{\rm CW}$ is the paramagnetic Curie temperature.
We recall that in the Weiss mean field theory, (-$\theta_{\rm CW}$) $\theta_{\rm CW}$ corresponds to the (anti-) ferromagnetic transition temperature.
However, this theory does not take into account single ion effects such as the crystal field which are expected to be significant in PuPd$_3$.

There is a field dependent residual susceptibility $\chi_0(H)$ which mainly arises from impurities, the encapsulation and the sample holder. In
addition, the Pauli susceptibility of the conduction electrons may also contribute to $\chi_0$ and can be estimated from the electronic Sommerfeld
coefficient of LuPd$_3$, $\gamma^+_{\mathrm{LuPd_3}}$=3.2(1)~mJmol$^{-1}$K$^{-2}$, which yields $\chi_{\mathrm{Pauli}} \approx 5.2(2) \times 10^{-5}
\mu_B$ T$^{-1}$ f.u.$^{-1}$. This is significantly lower than the observed values of the residual susceptibility, which are of the order of 
$10^{-3}$ $\mu_B$T$^{-1}$ f.u.$^{-1}$,
indicating that the conduction electron susceptibility contribution is negligible. The fitted parameters to the Curie-Weiss relation for each sample
are given in table~\ref{tab:pulupd3cpcwinfo}. The quoted error is deduced from calculating the covariance matrix of the parameters from the covariance 
matrix of the data~\cite{bard} assuming that this is diagonal and proportional to $\epsilon^{-2}$ where $\epsilon$ is the standard error in the
measured moment or heat capacity as determined by the \emph{MultiVu} software supplied by Quantum Design.
Figure~\ref{fg:pulupd3invchi} shows the inverse susceptibility of the different compositions with the residual susceptibility $\chi_0$ subtracted.

The magnitudes of the effective moments are all significantly higher than the $LS$-coupling value, 0.85~$\mu_B$. Any crystal field interaction will
only decrease this effective moment because as the crystal field split levels become further separated and hence thermally de-occupied, their angular
momentum will cease to contribute to the moment. The effective moment with zero crystal field splitting in \emph{intermediate coupling} on the other
hand is approximately 1.4~$\mu_B$, as calculated using the theory outlined in section~\ref{sec-numcalcs}. This value may be decreased slightly by a
large crystal field, and suggests that we should use intermediate coupling to calculate the single-ion properties of Pu$^{3+}$.

An alternative reason for the higher than expected effective moment may be due to some high moment paramagnetic impurity. However, analysis of the x-ray
diffraction patterns showed that the only detectable impurity is Lu$_2$O$_3$ which is non-magnetic. There may also be some trace amounts of oxides of Pu
which is not observed in the diffraction pattern. PuO$_2$ is a Van Vleck paramagnet~\cite{raphael_lallement} which may contribute to the impurity term
in equation~\ref{eq:curieweiss}, whilst Pu$_2$O$_3$ is an antiferromagnet with an effective moment of 2.1~$\mu_B$~\cite{mccart}. However, one would
need approximately 6 mol \% Pu$_2$O$_3$ in PuPd$_3$ order for it be responsible for the increased effective moment compared to the $LS$-coupling
expectation, at which concentration it should be detectable in the X-ray diffraction pattern, which is not the case. Furthermore, the enhanced value
of $\mu_{\mathrm{eff}}$ for Pu$_2$O$_3$ which also has a free ion Pu$^{3+}$ configuration suggests that intermediate coupling is appropriate in these
cases.

\begin{table*} \renewcommand{\arraystretch}{1.3}
\begin{center}
   \begin{tabular*}{15.5cm}{@{\extracolsep{\fill}}l|c|c|c|c|c|c|l} 
  Composition \qquad\qquad  &\mc{2}{c|}{$T_N$ (K)}&\mc{2}{c|}{$\gamma^{\pm}$(mJ mol$^{-1}$K$^{-2}$)}&$\theta_D$ (K)~~&$\mu_{\rm eff}$ ($\mu_B$ Pu$^{-1}$)&$\theta_{\rm CW}$ (K)
      \\\cline{2-5}
	                      &     $C_p$      &    $\chi$      &   $T>T_N$   &  $T<8$ K   &  $T>T_N$     &                   &                        \\ \hline
            PuPd$_3$          &~~~~ 24.4(3) ~  & ~  24.9(4)  ~~ &             & ~ 75(1) ~ ~& ~306(19)~~~  &  1.02(1)  ~~      &  -39.9(1) ~            \\
   Pu$_{0.9}$Lu$_{0.1}$Pd$_3$ &~~~~ 22.7(4) ~  & ~  22.5(1)  ~~ &             &  133(8) ~ ~& ~304(18)~~~  &  0.97(2)  ~~      &  -15.8(2) ~            \\
   Pu$_{0.8}$Lu$_{0.2}$Pd$_3$ &     20(1)      & ~  19.5(1)  ~~ &             &  221(14)  ~& ~280(6)~~~~~~&  0.97(2)  ~~      &   -7.5(2)              \\
   Pu$_{0.5}$Lu$_{0.5}$Pd$_3$ &     15(1)      & ~  12.8(1)  ~~ &    183(36)~~&  289(10)  ~& ~270(10)~~~  &  1.15(2)  ~~      &  ~ 1.8(1)              \\
   Pu$_{0.2}$Lu$_{0.8}$Pd$_3$ & ~~   7(2)      &                & ~~  99(49)~~&  170(5) ~ ~& ~257(13)~~~  &  1.25(1)  ~~      &   -7.2(2) ~            \\
            LuPd$_3$          &                &                &             &~~~~3.5(5) ~& ~291(12)~~~  &                   &                        \\
\hline
   \end{tabular*}
   \caption{Transition temperatures and other parameters derived from the magnetic susceptibility and heat capacity of Pu$_{1-x}$Lu$_x$Pd$_3$.} 
   \label{tab:pulupd3cpcwinfo}
\end{center}
\end{table*}

\subsection{Heat Capacity} \label{sec-cp}

\begin{figure}
  \begin{center}
    \includegraphics[width=0.95\columnwidth]{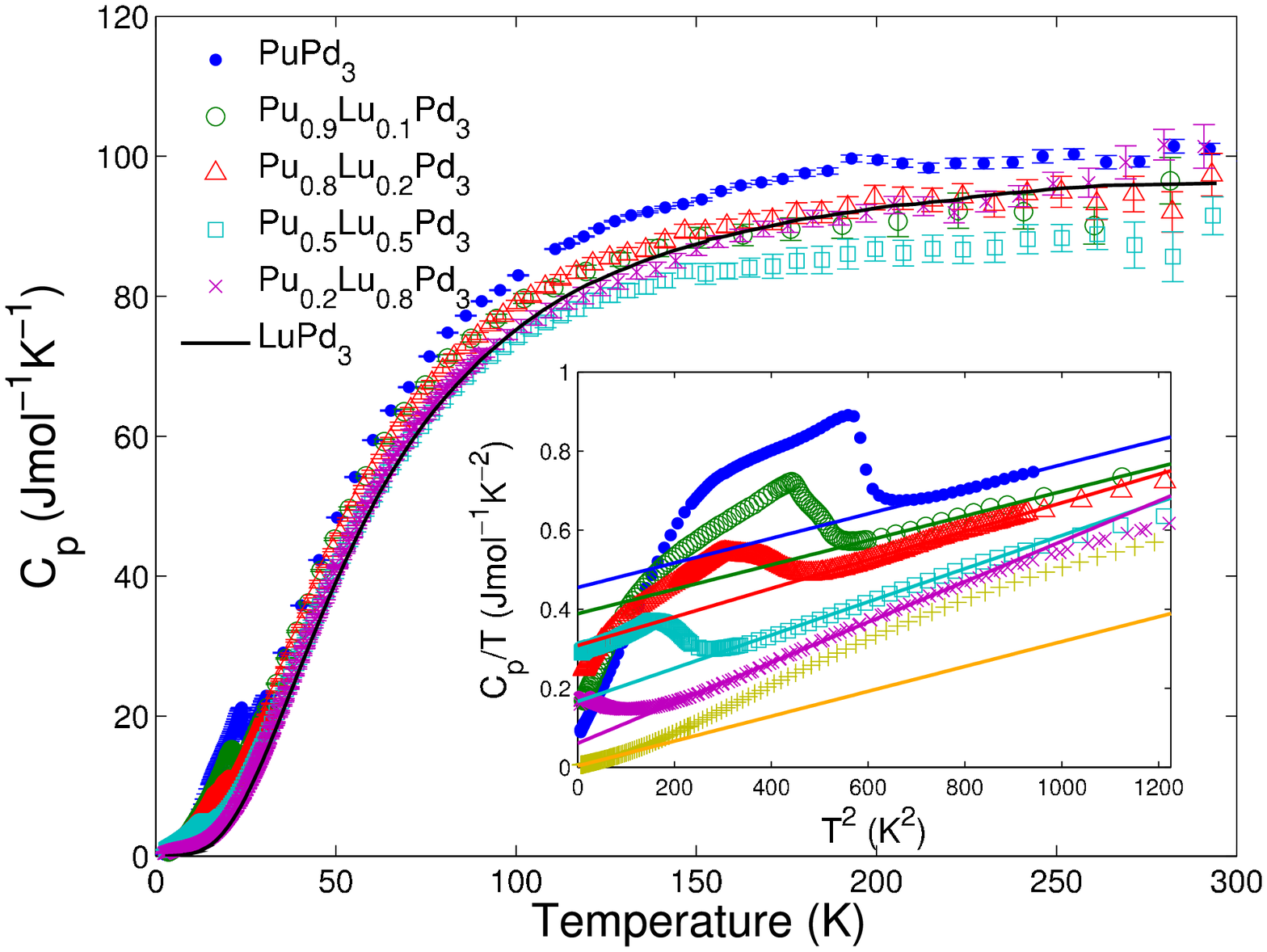}
    \caption{(Color online) Heat capacity of Pu$_{1-x}$Lu$_x$Pd$_3$ at zero applied magnetic field. The inset shows $C_p/T$ vs $T^2$ and solid lines there
     show fits to $C/T \sim \gamma + AT^2$. LuPd$_3$ data is shown in orange in the inset} \label{fg:pulupd3cp} 
  \end{center}
\end{figure}

\begin{figure}
  \begin{center}
    \includegraphics[width=0.95\columnwidth]{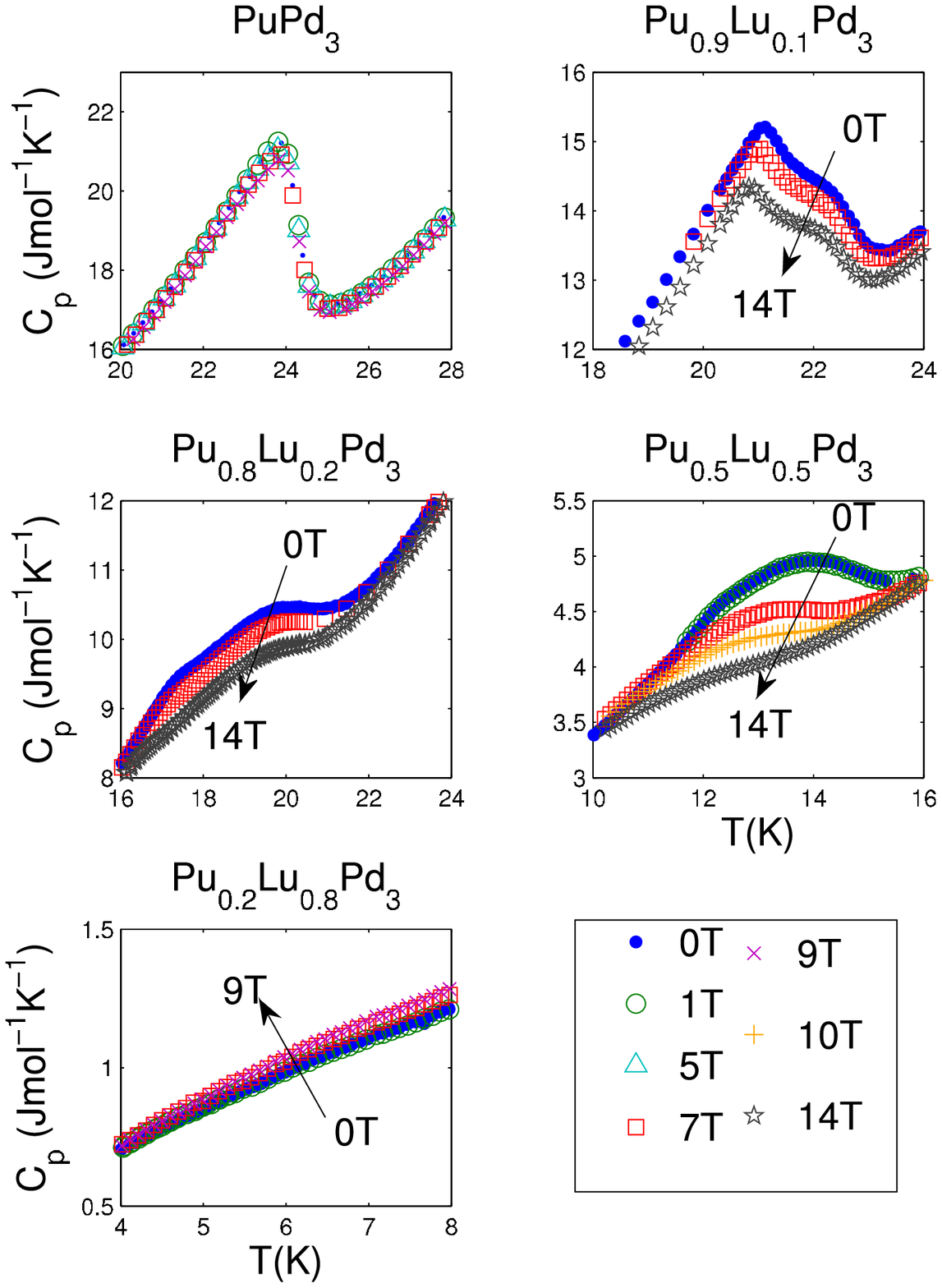}
    \caption{(Color online) Heat capacity of Pu$_{1-x}$Lu$_x$Pd$_3$ in applied magnetic field. The arrows indicate the direction of increasing field.} 
    \label{fg:pulupd3cpfield}
  \end{center}
\end{figure}

The heat capacity at zero field is shown in figure~\ref{fg:pulupd3cp}, whilst details of the results in applied fields up to 14~T are in
figure~\ref{fg:pulupd3cpfield}. We note that at high temperatures, $C_p$ tends to the classical Dulong-Petit limit, $3NR=99.8$~Jmol$^{-1}$K$^{-1}$.
For Pu$_{0.9}$Lu$_{0.1}$Pd$_3$, the derivative in the heat capacity shows two minima, which stem from the step-like nature of the transition. The
higher temperature inflexion point at $\sim$22~K corresponds well with the peak in the inverse susceptibility, but the lower temperature peak at
$\sim$21~K does not match any feature in the magnetic susceptibility. Nevertheless, these two anomalies raise the possibility that there may indeed be
two transitions in this compound. Moreover, as can be seen in Figure~\ref{fg:pulupd3cpfield}, the heat capacity of Pu$_{0.8}$Lu$_{0.2}$Pd$_3$ also
shows indications of two transitions. 

An estimate of the electronic specific heat $C_{\mathrm{el}} = \gamma T$ and Debye temperature $\theta_D$ was obtained using the approximation 

\begin{equation} \label{eq:cp}
C \sim \gamma T + \frac{12\pi^4 N_A k_B}{5} \left(\frac{T^3}{\theta_D^3}\right)
\end{equation}

\noindent which is valid at low temperatures ($T\ll\theta_D$), from a plot of $C_p/T$ vs $T^2$ shown as an inset in figure~\ref{fg:pulupd3cp}. 
However, the magnetic heat capacity complicates the determination of $\gamma$ because the N\'eel temperature is very low in some of the Lu-rich
compounds. This increases the low temperature $C_p$ and hence the estimate of $\gamma$ from the straight line intercept.
For this reason, for the Lu-rich compositions, we show the results of fitting the data in the region above $T_N$ ($\gamma^+$) in addition to that
below 8~K ($\gamma^-$) in table~\ref{tab:pulupd3cpcwinfo}. For Pu-rich compositions, the data above $T_N$ will be affected by the Schottky anomaly at
approximately 17~K, and will be unreliable. Thus it appears from these estimates that the electronic heat capacity first increases with increasing $x$
until $x\approx 0.5$, whereupon it falls as $x$ rises further.
The spread in the fitted parameters when data from different ranges of temperatures in the region $25<T<40$~K for $\gamma^+$, and $2<T<8$~K for
$\gamma^-$ was taken as an estimate of the errors in these parameters. 

%
%
%

\begin{figure} \begin{center}
    \includegraphics[width=0.95\columnwidth]{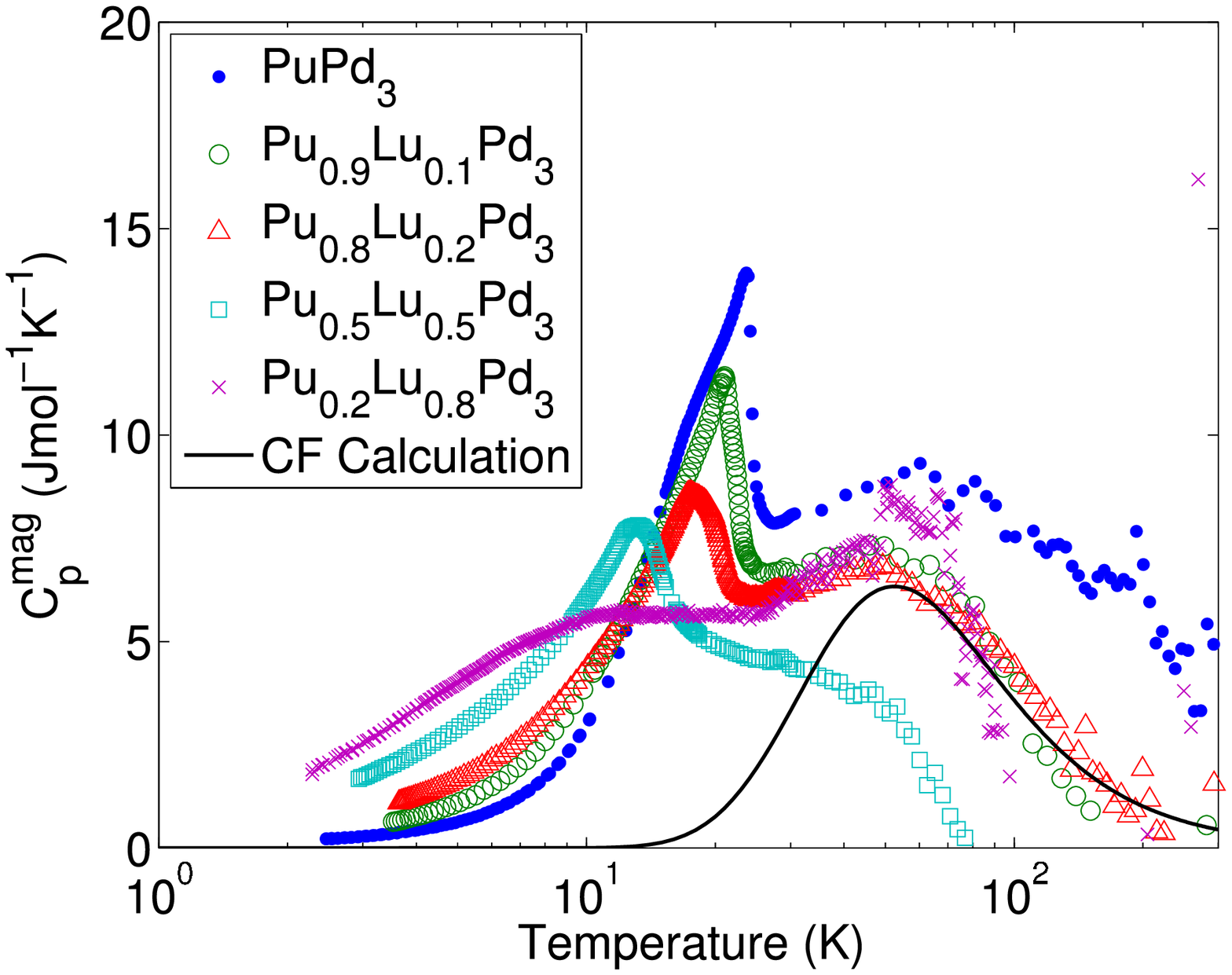}
    \caption{(Color online) Deduced magnetic and electronic contribution to the heat capacity {\rm$C_p^{\mathrm{mag}}=C_p$(Pu$_{1-x}$Lu$_x$Pd$_3$) -
    ($C_p$(LuPd$_3$)-$\gamma^+_{\rm Lu}T$)}. 
        The solid line is a crystal field calculation.} \label{fg:pulupd3cpmag}.
\end{center} \end{figure}

The Debye temperature $\theta_D$ was determined from fitting the high temperature data, and appears to be independent of Lu-doping. This
suggests that the phonon contribution to the heat capacity is constant through the series. A good estimate of this contribution is given by the heat
capacity of the non-magnetic isostructural compound LuPd$_3$, which also has a negligible electronic heat capacity,
$\gamma_{\mathrm{Lu}}=3.5(5)$~mJmol$^{-1}$K$^{-2}$. We have thus extracted the additional electronic \emph{and} magnetic heat capacity of
Pu$_{1-x}$Lu$_x$Pd$_3$ by subtracting that of LuPd$_3$, as

\begin{equation} \label{eq:cpmag}
C_p^{\mathrm{mag}}=C_p(\mathrm{Pu}_{1-x}\mathrm{Lu}_x\mathrm{Pd}_3)-(C_p(\mathrm{LuPd}_3)- \gamma^+_{\mathrm{Lu}}T)
\end{equation}

\noindent This extracted quantity, scaled by the Pu concentration, is shown in figure~\ref{fg:pulupd3cpmag}. 

The magnetic heat capacity for all the compounds shows a peak at $\sim$60~K, which we attribute to a Schottky anomaly from the crystal field (CF)
splitting. The cubic CF on the Pu$^{3+}$ ions splits the six-fold ground multiplet ($J=\frac{5}{2}$ in $LS$-coupling) into a doublet and quartet. The
energy gap, $\Delta^{\rm CF}$, between these two levels determines the temperature of the Schottky peak, such that $\Delta^{\rm CF}\sim$12~meV
corresponds to a peak at 60~K. The magnitude of this peak, however, is determined by whether the doublet ($C_p$=6.3~Jmol$^{-1}$K$^{-1}$) or quartet
($C_p$=2~Jmol$^{-1}$K$^{-1}$) is the ground state. The data in figure~\ref{fg:pulupd3cpmag} thus suggest a doublet ground state.

\begin{figure} \begin{center}
    \includegraphics[width=0.95\columnwidth]{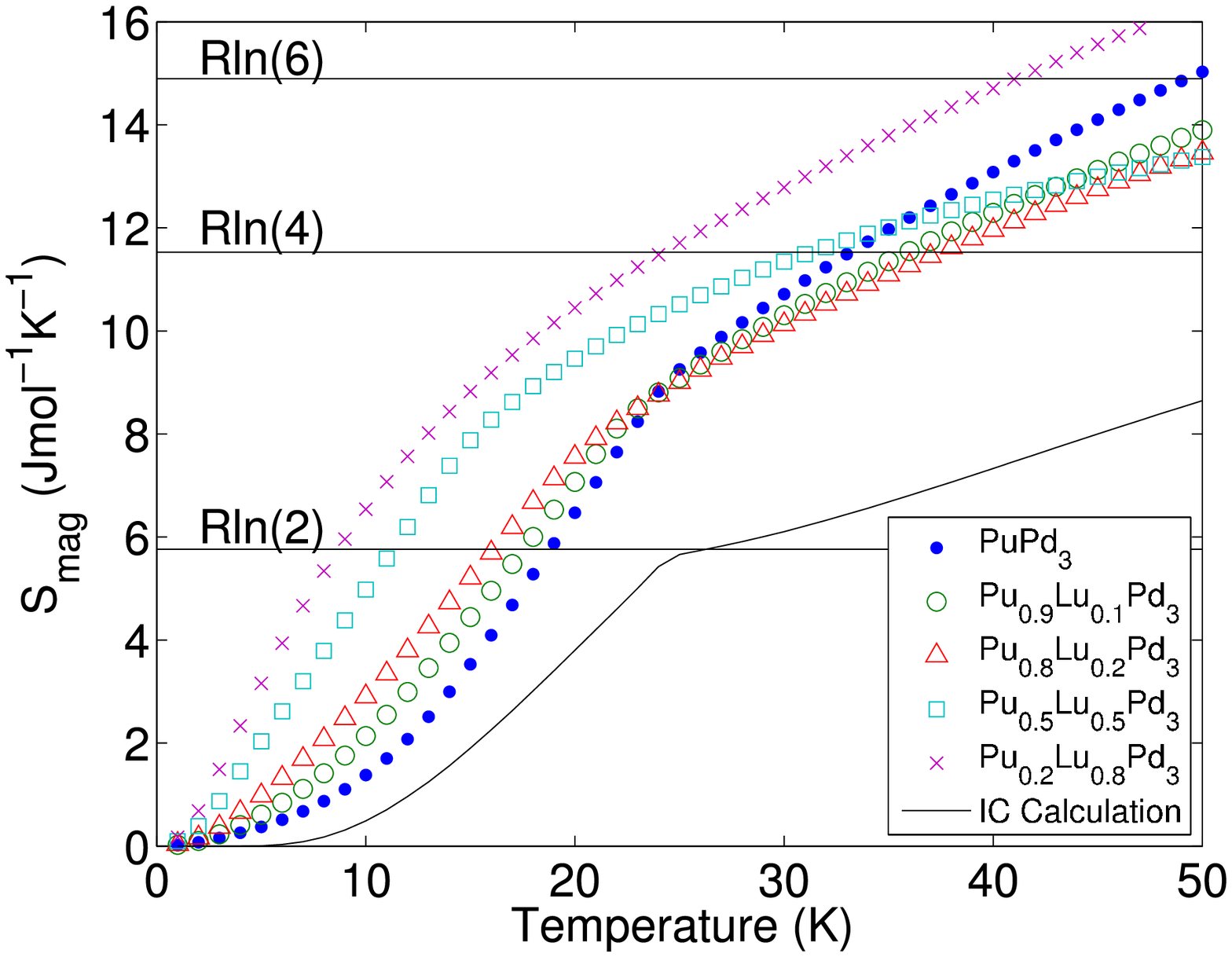}
    \caption{(Color online) Deduced magnetic contribution to the entropy of Pu$_{1-x}$Lu$_x$Pd$_3$, calculated by numerically integrating $C_p^{\mathrm{mag}}/T$.}
             \label{fg:pulupd3Smag}.
\end{center} \end{figure}

This is supported by the magnetic entropy, shown in figure~\ref{fg:pulupd3Smag}, deduced by numerically integrating the magnetic heat capacity, $S(T)
= \int_2^T \frac{C_p^{\mathrm{mag}}}{T} dT$. From the very low heat capacity of PuPd$_3$ at low temperatures, we believe the magnetic entropy from 0
to 2~K is negligble, and have not included this range in the integration. The value of the entropy at the N\'eel temperature is approximately $R\ln(3)$
for PuPd$_3$, which is above the value, $R\ln(2)$, expected for a doublet ground state. If the electronic heat capacity $\gamma^-_{\rm
Pu}=76$~mJmol$^{-1}$K$^{-2}$ is subtracted from the integral, then we obtain $S(T_N)\approx R\ln(2.4)$. The remaining discrepancy may be due to (i) an
incomplete subtraction of the phonon contribution, as the heat capacity of LuPd$_3$ may not be exactly analogous to the phonon heat capacity of
PuPd$_3$, and (ii) a larger value of $\gamma$ (see the discussion in section~\ref{sec-numcalcs}).

Finally, the inset to figure~\ref{fg:pulupd3cp} shows what appears to be a hump at $\sim$17~K in the heat capacity of PuPd$_3$. This feature was
initially attributed to the Schottky anomaly from the CF splitting in reference~\cite{le07pupd3}. However it is more likely due to a Schottky anomaly
from the splitting of the doublet ground state in the ordered phase, and indeed such a feature is observed in the mean field calculations in the next
section. A splitting of $\Delta^{\rm MF}=3.5$~meV gives a peak at $\sim$17~K, which is reasonable.

\section{Mean Field Calculations} \label{sec-numcalcs}

As noted in section~\ref{sec-pupd3-expMag}, the measured effective moment for both PuPd$_3$ and the doped compounds is approximately 1~$\mu_B$/Pu, in
contrast to the expected $LS$-coupling value of $g_J\sqrt{J(J+1)} = 0.85$~$\mu_B$ from a Hund's rule $^6$H$_{5/2}$ ground state for a Pu$^{3+}$ ion.
$LS$-coupling is a good approximation when both the Coulomb ($\mathcal{H}_{\mathrm C}$) and spin-orbit ($\mathcal{H}_{\mathrm so}$) interactions are
large but $\mathcal{H}_{\mathrm C} \gg \mathcal{H}_{\mathrm{so}}$. In contrast, when $\mathcal{H}_{\mathrm C} \ll \mathcal{H}_{\mathrm{so}}$,
$jj$-coupling is a better approximation, whereupon we obtain $\mu_{\mathrm{eff}}$ = 2.86~$\mu_B$. In between these limits, for the case of intermediate
coupling, the effective moment is a function of $\mathcal{H}_{\mathrm C}$ and $\mathcal{H}_{\mathrm{so}}$, and the full Hamiltonian, including both
these terms and the crystal field ($\mathcal{H}_{\mathrm{cf}}$) and Zeeman interactions ($\mathcal{H}_{\mathrm Z}$) must be calculated.

\begin{figure}
  \begin{center}
    \includegraphics[width=1.0\columnwidth]{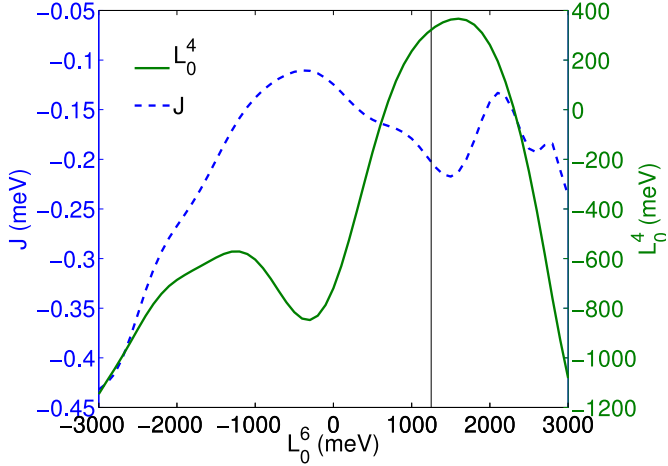}
    \caption{(Color online) Calculated dependence of the CF parameter $L_0^4$ and the nearest neighbour exchange parameter 
             $\mathcal{J}$ on $L_0^6$ subject to constraints described in the text. The vertical line indicates the optimal parameters.}
    \label{fig-b4b6_J}
  \end{center}
\end{figure}

\begin{figure}
  \begin{center}
    \includegraphics[width=0.95\columnwidth]{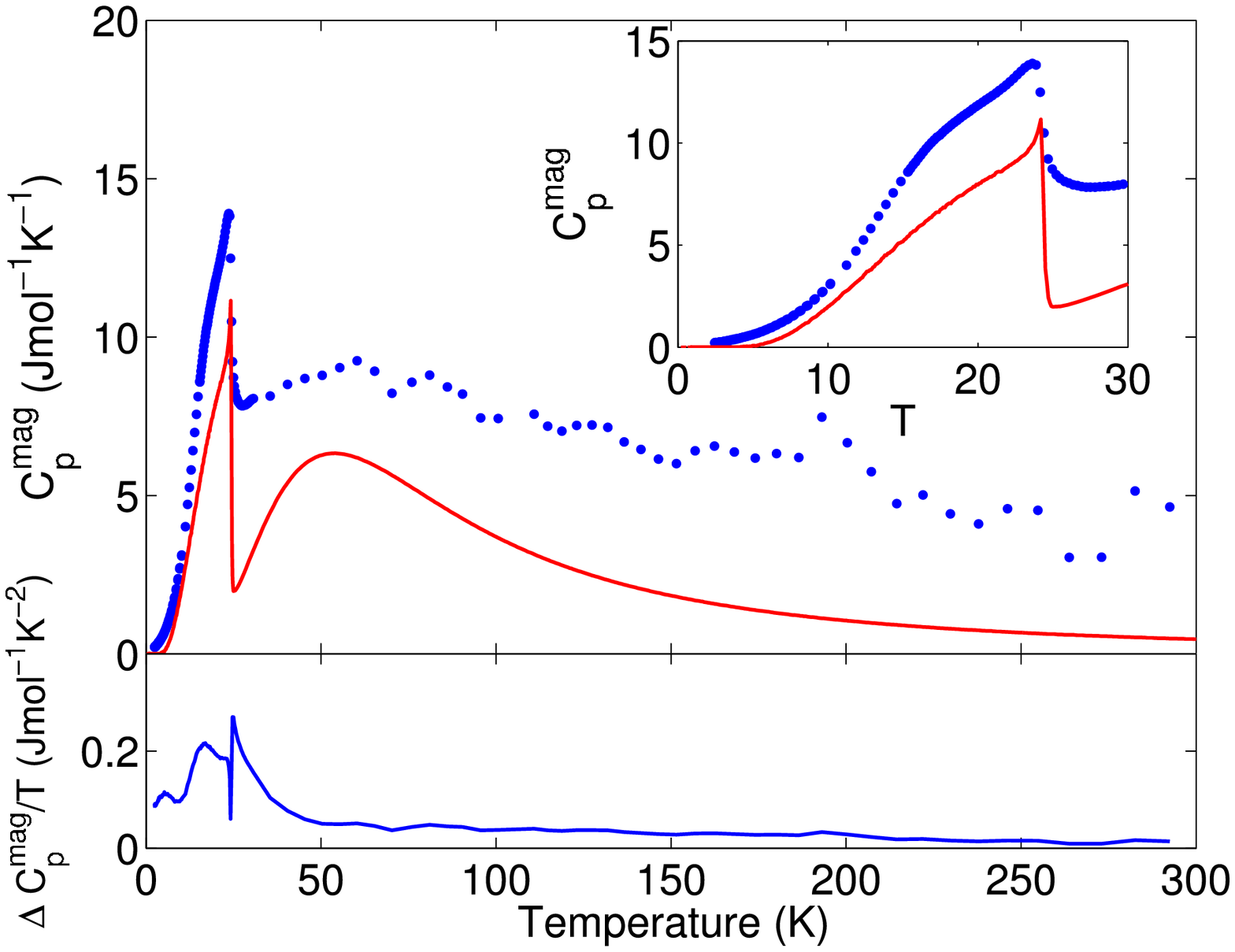}
    \caption{(Color online) Measured and calculated magnetic heat capacity of PuPd$_3$. The estimated electronic heat capacity, $\Delta C_p^{\mathrm{mag}}/T = \frac{C_p^{\mathrm{mag}} 
     - C_p^{\mathrm{ic}}}{T}$ is given in the bottom panel and has a mean value of 122~mJmol$^{-1}$K$^{-2}$.}
    \label{fig-mfcp}
  \end{center}
\end{figure}

The strength of $\mathcal{H}_{\mathrm C}$ and $\mathcal{H}_{\mathrm{so}}$, parameterised by the Slater ($F^k$) and spin-orbit ($\xi$) integrals, is
fixed by the atomic environment of the unfilled shell electrons. Thus in practice, intermediate coupling refers to the case where the value of
$F^k$ and $\xi$ are determined either from ab-initio (Hartree-Fock) calculations, or from experimental measurements using optical spectroscopy. Using
parameters determined experimentally by Carnall from the spectra of dilute Pu$^{3+}$ in LaCl$_3$~\cite{carnallAn}, the effective moment is 1.44~$\mu_B$. 
It is conceivable that in a metallic system like Pu$_{1-x}$Lu$_x$Pd$_3$ there may be small changes to $F^k$ and $\xi$ compared to the insulating
salts on which the measurements of~\cite{carnallAn} were made~\cite{taylorPrHET}. Nevertheless, a 10~\% change in $F^k$ and $\xi$ to make the system
more $LS$-like only yields $\mu_{\mathrm{eff}}=$1.38~$\mu_B$. In order to obtain $\mu_{\mathrm{eff}}\sim$1~$\mu_B$, we must double the magnitude of
$F^k$ and $\xi$ compared to their measured values, which is probably unphysical.

Given that the crystal field interaction is small, as judged by the $\sim$12~meV split between the doublet ground state and first excited quartet
deduced from the heat capacity measurements, $\mathcal{H}_{\mathrm{cf}}$ has little effect on $\mu_{\mathrm{eff}}$. Thus we believe that the lower
than expected effective moment is most likely due to Kondo screening. Nonetheless, a mean field calculation which can model the antiferromagnetic
order and the single ion intermediate coupling behaviour is still valuable to interpret the heat capacity and magnetisation data. Such a calculation,
carried out using the {\sl McPhase} package~\cite{rotter04}, is detailed below. 

We have assumed a nearest neighbour only exchange interaction between 5$f$ electrons, which is reasonable given the G-type antiferromagnetic structure
where nearest neighbour moments align in antiparallel. Thus, there are three free parameters in the calculations: two crystal field (CF) parameters,
$L_0^4$ and $L_0^6$, and one exchange parameter $\mathcal{J}$. There are two other non-zero CF parameters, but they are fixed by the cubic point
symmetry of the Pu$^{3+}$ ions such that $L_4^4=\sqrt{5/14}L_0^4$ and $L_4^6=-\sqrt{7/2}L_0^6$. It should be noted that the parameters used here
correspond to the \emph{Wybourne} normalisation~\cite{NN00}, rather than the usual \emph{Stevens} normalisation (usually denoted $B_l^m$). This is
because the Stevens operator equivalents $O_l^m$ are valid only within a single multiplet of given $J$, whereas we now require operators that can span
all the allowed $J$ values.

The two CF parameters are fixed by the requirement that they result in a doublet ground state with a quartet at $\sim$12~meV. This fixes a relation
between $L_0^4$ and $L_0^6$ as shown in figure~\ref{fig-b4b6_J}. The N\'eel temperature $T_N$ then fixes a relation between $\mathcal{J}$ and the crystal
field parameters, and finally the magnetisation below $T_N$ was used to fixed all three values, yielding $L_0^4=320$, $L_0^6=1250$, and
$\mathcal{J}=-0.204$ meV.

The magnetisation is calculated by including in the Hamiltonian a Zeeman term, $-\mu_B~(\V{L}+2\V{S})\cdot\V{H}$; numerically diagonalising the energy
matrix and calculating the expectation value of the moment operator $\V{L}+2\V{S}$. The calculated inverse susceptibility is shown as a solid black
line in figure~\ref{fg:pulupd3invchi} for comparison with the measured data. Unfortunately, better agreement with the data within the constraints of
the mean-field intermediate coupling model is only possible by increasing the Coulomb or spin-orbit integrals to unphysical values. A more likely
explanation is the suppression of the effective moment by Kondo screening, which is not considered in the current model.

The heat capacity is calculated by numerically differentiating the internal energy, $\langle U\rangle=\sum_n E_n\exp(-E_n/k_BT)/Z$, by the
temperature, and the entropy by subsequently numerically integrating this. The calculated heat capacity, shown in figure~\ref{fig-mfcp}, shows a
shoulder around $\sim$20~K in accordance with the data which arises from a Schottky peak due to the splitting of the ground state doublet in the
ordered phase. We can also estimate the electronic heat capacity by subtracting this calculated $C_p^{\mathrm{ic}}$ from the measured electronic and
magnetic heat capacity, $C_p^{\mathrm{mag}}$, the result of which is shown in the bottom panel of figure~\ref{fig-mfcp}. The spike near $T_N$ is due
to the differences in the sharpness of the calculated and measured transitions in this temperature range. Overall however, the mean value,
$\bar{\gamma}=$122~mJmol$^{-1}$K$^{-2}$, over the full temperature range is in fair agreement with that derived from the low temperature heat
capacity, $\gamma^-_{\rm Pu}$=76~mJmol$^{-1}$K$^{-2}$.

Similar mean-field heat capacity calculations for the other compositions, where the exchange coupling $\mathcal{J}$ was reduced to reflect the lower
$T_N$, did not yield the broad transitions seen in figure~\ref{fg:pulupd3cpmag}, but rather the sharp lambda anomalies expected of an
antiferromagnetic transition. Thus a subtraction to deduce their electronic specific heat becomes increasingly untenable. The broad transitions
observed in the data are probably due to disorder in the system as a result of the Lu doping.

Finally, the calculated internal fields in the model are 226~T (180~T) at 1~K (20~K), which agree well with the molecular field of 217~T determined
from fitting the measured resistivity using a simple CF model, as described in the next section.

\section{Electrical Transport Measurements} \label{sec-elec}

Electrical transport measurements were carried out using thin parallel-sided samples extracted after crushing the polycrystalline buttons produced by
arc-melting. The first transport measurements were completed some months after the production of the samples, so there were significant aging effects
in the Pu samples. This prompted us to anneal the PuPd$_3$ sample, and re-measure its resistivity, resulting in a large decrease in the residual
resistivity $\rho_0$ from 225 $\mu\Omega$ cm to 11 $\mu\Omega$ cm. Subsequently the resistivity of the other compositions was also re-measured after
annealing. The values of $\rho_0$ show a rapid increase with Lu doping up to $x=0.5$ thereafter decreasing with $x$, as summarised in
table~\ref{tab:pulupd3cfres_pars}. This is due to the increasing number of defects caused by Lu substitution. Finally, measurements of the Hall
coefficient and longitudinal resistivity of PuPd$_3$ and the $x$=0.1,0.2,0.5 compositions in field were also carried out.

\begin{figure}
  \begin{center}
    \includegraphics[width=0.95\columnwidth]{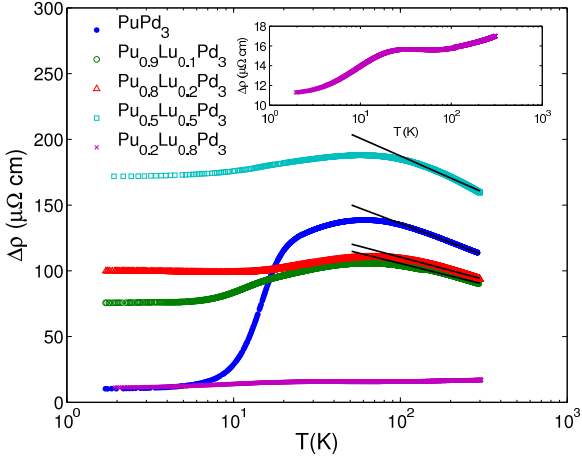}
    \caption{(Color online) Zero-field magnetic resistivity of (Pu,Lu)Pd$_3$ on a logarithmic scale. Inset is an enlarged view of the 
             Pu$_{0.2}$Lu$_{0.8}$Pd$_3$ composition. Solid lines are fits to the relation $\rho_0-\rho_1\log(T)$ described in the text.}
    \label{fg:pulupd3logres}
  \end{center}
\end{figure}

The resistivity of LuPd$_3$ is well fitted by the Bloch-Gr\"uneisen relation 

\begin{equation} \label{eq:bloch-grueneisen}
\rho=C\frac{T^5}{\theta_D^6}\int_0^x \frac{x^5}{(e^x-1)(1-e^{-x})} dx
\end{equation}

\noindent where $x=T/\theta_D$, with $C=181(1)$~$\mu\Omega$ cm and $\theta_D=166(1)$~K. It was taken to be representative of the non-magnetic
contribution to the resistivity of Pu$_{1-x}$Lu$_x$Pd$_3$, and used to estimate the magnetic resistivity as $\Delta\rho =
\rho$(Pu$_{1-x}$Lu$_x$Pd$_3$)$-\rho$(LuPd$_3$). This quantity is plotted in the case of zero applied magnetic field in figure~\ref{fg:pulupd3logres}.
The in-field measurements showed little change from the zero field data, and the data for PuPd$_3$ agree well with previous
measurements~\cite{harvey73}, albeit with a slightly lower residual resistivity. 

Qualitatively, the behaviour of the resistivity may be divided into a high temperature Kondo-like regime, where the resistivity increases with decreasing
temperature until $\sim$50~K, followed by the onset of coherence, from where it falls sharply with temperature, and shows no clear anomaly at $T_N$. 
At low temperatures, the resistivity follows an exponential temperature behaviour, in contrast the power law behaviour expected in metals from the
Bloch-Gr\"uneisen relation. Electrons scattering from antiferromagnetic magnons will give rise to an exponential temperature dependence, as will
spin-disorder scattering from the localised $5f$ moments themselves. A fit to the electron-magnon resistivity~\cite{nhandersen} assuming an isotropic
magnon dispersion $\omega=\sqrt{\Delta^2+Dk^2}$, such that~\cite{gofryk_pupd5al2}

\begin{eqnarray} \label{eq:afmres}
\rho_{\mathrm e-m} &=& \rho_0 + B\Delta^2\sqrt{\frac{k_BT}{\Delta}} e^{-\Delta/k_BT} \\ \nonumber
                   & &\quad \times \left[ 1 + \frac{3\Delta}{2k_BT} + \frac{2}{15} \left( \frac{\Delta}{k_BT} \right)^2 \right]
\end{eqnarray}

\noindent yields, however, a spin-gap $\Delta=4$~meV which is significantly larger than that expected from the heat capacity below $T_N$,
if the shoulder at 17~K corresponds to a Schottky peak which arises from a gap of approximately 3.2~meV. In contrast, this very splitting 
between the doublet ground states in the ordered phase is predicted by the spin-disorder resistivity model described below.

Above $\sim$70~K, the resistivity is
well fitted by a $\rho_0 - \rho_1\log(T)$ term~\cite{kondores}, where $\rho_0$ is the residual resistivity, and $\rho_1$ is proportional to the
interaction between the conduction electrons and Kondo impurities. The fit is shown in figure~\ref{fg:pulupd3logres},
with parameters $\rho_0 = 235.8(3)$ $\mu\Omega$ cm and $\rho_1 = 22(1)$ $\mu\Omega$ cm for PuPd$_3$. The value of $\rho_1$ initially decreases with Lu doping
to 15(2)~$\mu\Omega$ cm for $x=0.1$ and 16(2)~$\mu\Omega$ cm for $x=0.2$ but then increases to 27(3)~$\mu\Omega$ cm for $x=0.5$. This increase
suggests that the Kondo interaction is strengthened at half doping.

The magnetic resistivity of Pu$_{0.2}$Lu$_{0.8}$Pd$_3$ does not show the Kondo behaviour of the other compositions, but rather increases with
increasing temperature with a plateau region around 30-80~K. This behaviour and also the exponential temperature dependence of the low temperature
part of the resistivity 
is characteristic of a simple crystal field spin-disorder resistivity
model~\cite{raowallace}. This model is based on the scattering of conduction electrons with spin $\V{s}$ by a localised moment $\V{J}$ through an
exchange interaction $-2G\V{s}\cdot\V{J}$, giving the resistivity in the first Born approximation as

\begin{multline} \label{eq:cfres}
\rho_{\mathrm{s-f}} = \frac{3\pi Nm^*}{\hbar e^2 E_F} G^2 (g-1)^2 \frac{1}{Z} \\
   \times \sum_{m_s,m'_s,i,i'} \bra{m'_s,\psi'_i} \V{s}\cdot\V{J} \ket{m_s,\psi_i} p_i f_{ii'}
\end{multline}

\noindent where the occupation factor for the crystal field level at $E_i$ is $p_i = \exp(-\sfrac{E_i}{k_B T})$ and the conduction electron population
factor is $ f_{ii'} = 2[1+\exp(-\sfrac{E_i-E_{i'}}{k_B T})]^{-1}$. The wavefunctions $\ket{\psi_i}$ and energies $E_i$ are determined by diagonalising
the crystal field Hamiltonian.

In the absence of a crystal field, the $2J+1$ degenerate spin orbit ground state levels yield a temperature independent resistivity given by

\begin{equation} \label{eq:cfres0}
\rho^{(0)}_{\mathrm{s-f}} = \frac{3\pi Nm^*}{\hbar e^2 E_F} G^2 (g-1)^2 J(J+1)
\end{equation}

\noindent In our case, with a $J=\sfrac{5}{2}$ multiplet in a cubic crystal field, and in the absence of a magnetic field, equation~(\ref{eq:cfres})
reduces to a sum of exponential functions, because the wavefunctions $\ket{\psi_i}$ are fixed, and the crystal field parameter can only change the
splitting $\Delta^{\rm CF}$ between the quartet and doublet. There is thus a universal behaviour, with the resistivity tending to
$\rho^{(0)}_{\mathrm{s-f}}$ at $T \gg \Delta^{\rm CF}$, and then falling exponentially as the temperature falls below some level such that the excited
crystal field states are no longer populated. This temperature is approximately $0.4\Delta^{\rm CF}$, so the model suggests a splitting $\Delta^{\rm
CF} \approx 3.2$ meV, as the drop off in resistivity occurs around 15 K. This is significantly smaller than the splitting deduced from the Schottky
peak at $\sim$60~K but is similar to the splitting of the ground state doublet ($\Delta^{\rm MF}$) in the ordered phase, as determined by 
the shoulder at $\sim$17~K in the heat capacity data. 

\begin{table} \renewcommand{\arraystretch}{1.3}
\begin{center}
   \begin{tabular*}{8cm}{@{\extracolsep{\fill}}l|c|c|c||c} 
          Composition                &    $\rho_0$      & $\rho^{(0)}_{\mathrm{s-f}}$& $B_{\mathrm{mf}}$ & $B_{\mathrm{mf}}^{\mathrm{ic}}$  \\
                                     & ($\mu\Omega$ cm) &     ($\mu\Omega$ cm)       &       (T)         &           (T)                    \\ \hline
             PuPd$_3$                & ~~~   12.0(1)  ~ &       876(37)      ~       &     217(23)       &   ~       226                    \\
      Pu$_{0.9}$Lu$_{0.1}$Pd$_3$     &   ~   75(1)   ~~ &       118(5)       ~~      &      88(4)        &   ~       149                    \\
      Pu$_{0.8}$Lu$_{0.2}$Pd$_3$     &   ~   97(1)   ~~ &  ~     35(1)       ~~      &     213(6)   ~    &   ~       143                    \\
      Pu$_{0.5}$Lu$_{0.5}$Pd$_3$     &      172(1)  ~~~ &  ~     53(2)       ~       &      92(5)        &   ~       129                    \\
      Pu$_{0.2}$Lu$_{0.8}$Pd$_3$     & ~~~   11.5(1)  ~ &  ~~    17.8(1)             &      51(1)        &   ~ ~      88                    \\
   \hline
   \end{tabular*}
   \caption{Parameters for the spin-disorder resistivity model for Pu$_{1-x}$Lu$_x$Pd$_3$ described in the text. The crystal field parameter 
                  $B_4$=0.041(1) meV was determined using the data for PuPd$_3$, and thereafter fixed in the fitting of the other compositions.}
   \label{tab:pulupd3cfres_pars}
\end{center}
\end{table}

In order to accommodate this splitting, we introduce a molecular field $B_{\mathrm{mf}}$ term. The low temperature exponential increase is then governed
primarily by the split doublet, with a second exponential step at higher temperatures due to the crystal field splitting. It turns out that this
second step is not observed in the case of PuPd$_3$ because the two steps merge into each other. Indeed a fit to the data below $T_N$ with all
parameters in the spin-disorder resistivity model varying freely yields a CF splitting of 14~meV, in agreement with the heat capacity data. As $B_{\mathrm{mf}}$
decreases in line with $T_N$ with increasing Lu doping, whilst the CF splitting remains constant, the two steps become more pronounced in the
calculations. These two steps are observed in the case of Pu$_{0.2}$Lu$_{0.8}$Pd$_3$, but the second step is masked by the Kondo screening in the
other compositions.

\begin{figure}
  \begin{center}
    \includegraphics[width=0.9\columnwidth]{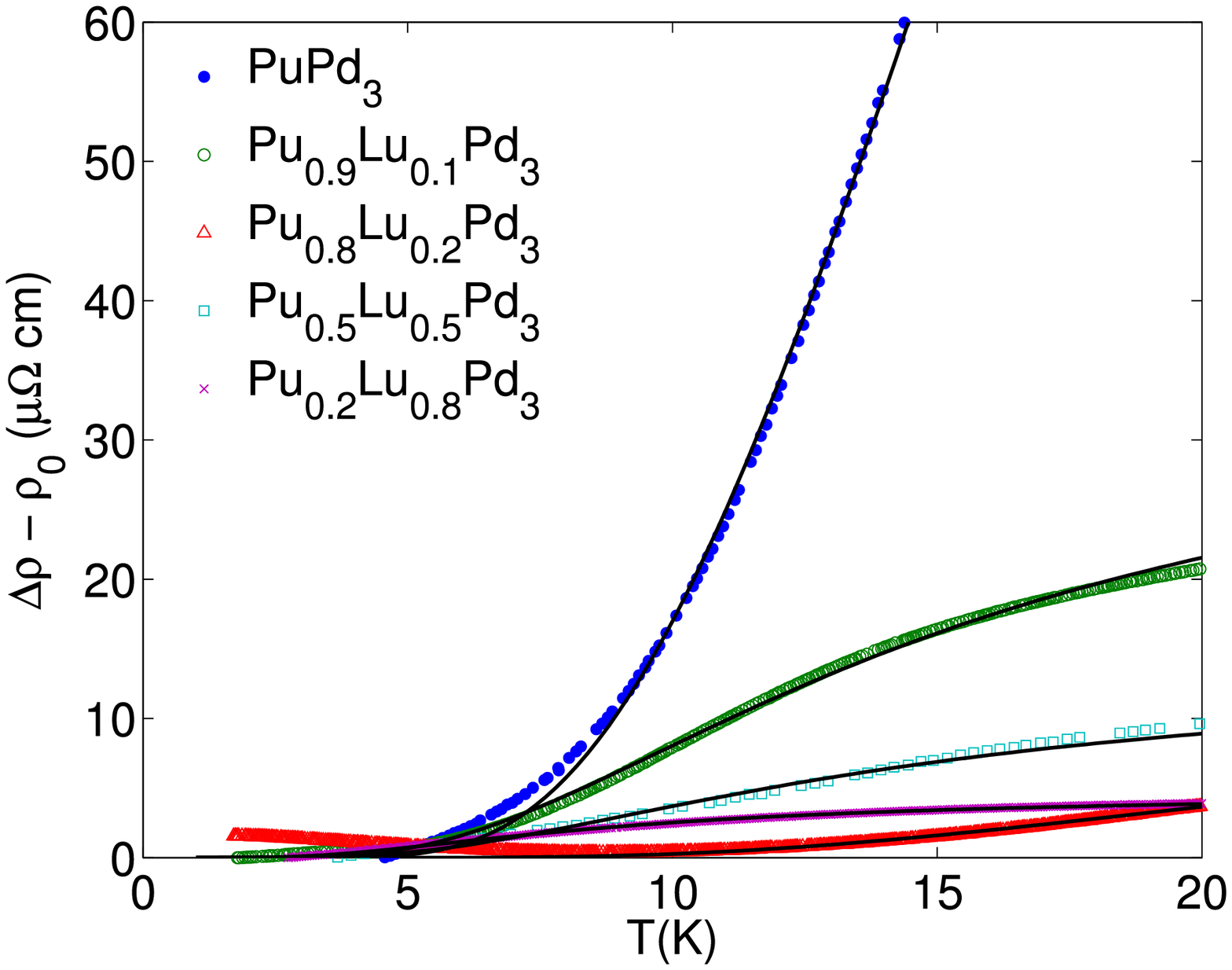}
    \caption{(Color online) Zero-field magnetic resistivity of (Pu,Lu)Pd$_3$ with residual resistivity $\rho_0$ subtracted. Solid lines are fits using the crystal
             field model described in the text} \label{fg:pulupd3cfres}
  \end{center}
\end{figure}

Using this simple crystal field model we obtained the parameters shown in table~\ref{tab:pulupd3cfres_pars}. Figure~\ref{fg:pulupd3cfres} 
shows the resulting fit to the data with $\rho_0$ subtracted. 
The CF splitting was fixed for all Lu doped compositions to the value determined from fitting the PuPd$_3$ data.
The molecular field determined from mean field calculations, $B_{\mathrm{mf}}^{\mathrm{ic}}$, with exchange parameters $\mathcal{J}=$ -0.202, -0.199, 
-0.1925, and -0.174~meV, for compositions $x$=0.1,0.2,0.5 and 0.8 respectively, is also shown in the table. Apart from the case of 
Pu$_{0.8}$Lu$_{0.2}$Pd$_3$, the fitted $B_{\mathrm{mf}}$ is consistently lower than the calculated $B_{\mathrm{mf}}^{\mathrm{ic}}$. This is due to the
overestimation of $T_N$ in the mean field approximation, because we have estimated $\mathcal{J}$ from $T_N$, and $B_{\mathrm{mf}}^{\mathrm{ic}}$ is
proportional to $\mathcal{J}$. Thus, $B_{\mathrm{mf}}^{\mathrm{ic}}$ is also overestimated. 

Pu$_{0.8}$Lu$_{0.2}$Pd$_3$ shows an upturn at low temperatures, which cannot be accounted for by the current model. In addition, this upturn affects
the fit by decreasing the ratio between the maximum and minimum resistivity, and hence $\rho^{(0)}_{\mathrm{s-f}}$. The exponential increase in the
resistivity also occurs at a higher temperature and over a broader temperature range in this composition than in the others, which explains the
anomalously high $B_{\mathrm{mf}}$. These features may be artefacts of the sample, because whilst an aged Pu$_{0.9}$Lu$_{0.1}$Pd$_3$ showed the upturn
at low temperatures, the annealed sample did not, whereas both aged and annealed Pu$_{0.8}$Lu$_{0.2}$Pd$_3$ samples showed the upturn. Furthermore the
resistivity of the aged Pu$_{0.8}$Lu$_{0.2}$Pd$_3$ sample is lower than that of the annealed sample. This suggests that the annealing had not fully repaired
the radiation damage, and thus the resistivity may be strongly affected by crystallographic defects.

Nevertheless when the fits were repeated using the calculated $B_{\mathrm{mf}}^{\mathrm{ic}}$ as fixed parameters, the fitted
$\rho^{(0)}_{\mathrm{s-f}}$ changed by less than 10\%. The fits thus showed that the $f$-conduction electron interaction decreases with Lu doping, with 
a slight increase for the $x=0.5$ composition compared to $x=0.2$ and $x=0.8$, in agreement with the fits to the Kondo parameter $\rho_1$.

\begin{figure}
  \begin{center}
    \includegraphics[width=0.9\columnwidth]{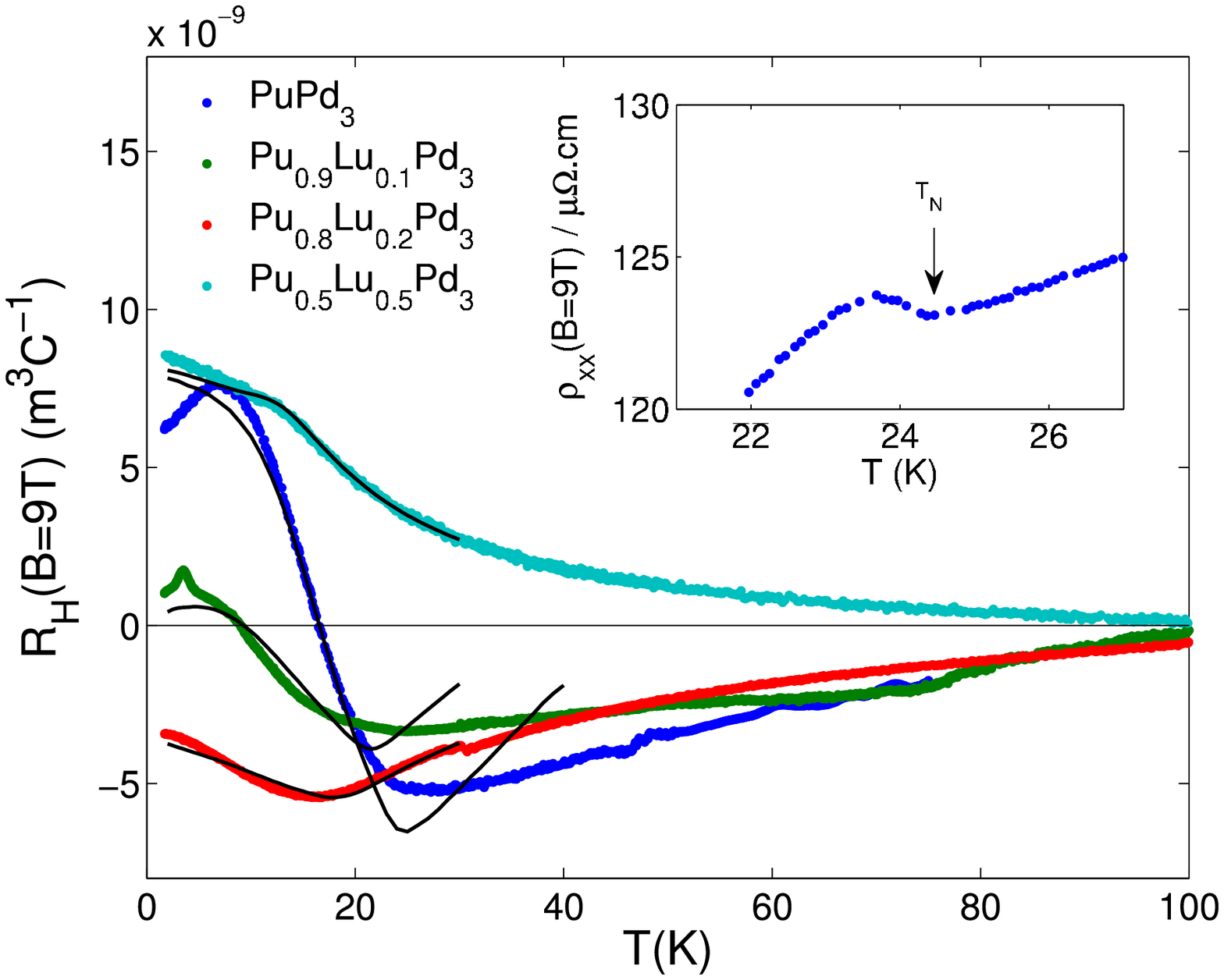}
    \caption{(Color online) Temperature dependence of the Hall coefficient $R_H = \rho_{xy}/B$ and magnetoresistance at 9~T (inset). The solid black line is a fit
                   using the measured magnetic susceptibility as described in the text}.
    \label{fg:pupd3hall}
  \end{center}
\end{figure}

\begin{table} \renewcommand{\arraystretch}{1.3}
\begin{center}
   \begin{tabular*}{8cm}{@{\extracolsep{\fill}}l|c|c|c|c} 
   $x=$   &     0         &     0.1       &     0.2       &    0.5         \\ \hline
   $R_0$  &    32.9(4)  ~ &    12.1(2)  ~ &     1.5(1)  ~ &    0.9(1)   ~~ \\
   $R_1$  &    -6.97(8)   &    -2.31(3)   &    -0.74(1)   &   -0.028(1) 
   \end{tabular*}
   \caption{(Color online) Fitted Hall constants of Pu$_{1-x}$Lu$_x$Pd$_3$. All values are in $(10^{-9}$m$^3$C$^{-1})$.}
   \label{tab:pulupd3hall_pars}
\end{center}
\end{table}

We now turn to the electrical transport properties in an applied magnetic field. The temperature dependence of the Hall coefficient $R_H$ and the
magnetoresistivity $\rho_{xx}$ in 9T are shown in figure~\ref{fg:pupd3hall}. $\rho_{xx}(T)$ follows the same behaviour as the zero field resistivity shown
above, with the exception that there is a small peak just below the N\'eel temperature as shown in the inset to the figure. This was not observed
previously and is reminiscent of the superzone scattering near $T_N$ in the heavy rare earths~\cite{mackintosh63}.

The temperature dependence of the Hall effect may be described phenomenologically by a scaling of the magnetisation, as in

\begin{equation} \label{eq:hallres}
\rho_{xy}(T) = R_0 B + R_1 \mu_0 M(T) 
\end{equation}

\noindent where $R_0$ is the \emph{ordinary} and $R_1$ the \emph{extraordinary} Hall constant. In figure~\ref{fg:pupd3hall}, the solid lines show fits
of the Hall coefficient to this relation using the measured magnetisation data. Both the ordinary and extraordinary Hall constants were found to
decrease with Lu doping, as shown in table~\ref{tab:pulupd3hall_pars}. $R_1$ is proportional to the conduction-$f$ electron exchange interaction
strength $G$ discussed above, so the decrease in its magnitude further indicates that this interaction becomes weaker with Lu doping.

Finally, we observed that the magnetoresistance, $\rho(B)$, is linear with applied field for all samples, and showed a negative slope at high
temperatures and a positive slope at low temperatures, except for LuPd$_3$ where the slope was always positive. We interpret the negative
magnetoresistance to be a sign of the Kondo effect at high temperatures, with the normal metallic behaviour giving a positive magnetoresistance at low
temperatures.

\section{Conclusions} \label{sec-pupd3-conc}

We have completed extensive bulk properties measurements on antiferromagnetic PuPd$_3$ and the pseudo-binary compounds Pu$_{1-x}$Lu$_x$Pd$_3$. The transition
temperature was found to decrease linearly from $T_N = $24.4(3)~K in PuPd$_3$ to 7(2)~K in Pu$_{0.2}$Lu$_{0.8}$Pd$_3$.

Heat capacity measurements show a Schottky anomaly at $\sim$60~K, which was interpreted as arising from a crystal field splitting between a doublet
ground state and an excited state quartet at $\sim$12~meV. The deduced Sommerfeld coefficient, $\gamma$ was found to be significantly higher than that
expected for the free electron model, with a value of the order of 0.1~Jmol$^{-1}$K$^{-2}$ determined by fitting the data directly and by subtracting
the calculated magnetic and measured phonon contributions. Direct fits to the data suggest that $\gamma$ decreases with increasing Lu substitution.
The magnetic heat capacity was calculated using a mean field model which showed that the shoulder in the data corresponds to a splitting of the
doublet ground state in the ordered phase with a gap of $\sim$3.5~meV. The size of this gap is supported by fits of the resistivity to a crystal field model.

Magnetic susceptibility and magnetisation measurements also showed that the paramagnetic effective moment, $\mu_{\mathrm{eff}}$ increases with Lu
concentration, approaching the value expected in intermediate coupling. These observations arise from the Kondo interaction
which suppresses the effective magnetic moment but enhances the electronic effective mass, $m^*$. As the Kondo interaction decreases
with Lu doping, $\mu_{\mathrm{eff}}$ is screened less, and $\gamma \propto m^*$ falls.

Electrical transport measurements support this decrease in the Kondo interaction with increasing Lu concentration, $x$, as parameters proportional to the
$f$-conduction electron coupling in a crystal field model of the resistivity and Hall effect were found to fall as $x$ increases.

\section*{Acknowledgements}

We thank G.H. Lander and K. Gofryk for helpful discussions.
M.D.L thanks the UK Engineering and Physical Sciences Research Council for a research studentship, and the Actinide User Lab at ITU. 
We are grateful for the financial support to users provided by the European Commission, Joint Research Centre within its "Actinide User Laboratory" program, and the
European Community's Access to Research Infrastructures action of the Improving Human Potential Programme (IHP), Contracts No. HPRI-CT-2001-00118, and
No. RITA-CT-2006-026176. 


\begin{thebibliography}{18}
\expandafter\ifx\csname natexlab\endcsname\relax\def\natexlab#1{#1}\fi
\expandafter\ifx\csname bibnamefont\endcsname\relax
  \def\bibnamefont#1{#1}\fi
\expandafter\ifx\csname bibfnamefont\endcsname\relax
  \def\bibfnamefont#1{#1}\fi
\expandafter\ifx\csname citenamefont\endcsname\relax
  \def\citenamefont#1{#1}\fi
\expandafter\ifx\csname url\endcsname\relax
  \def\url#1{\texttt{#1}}\fi
\expandafter\ifx\csname urlprefix\endcsname\relax\def\urlprefix{URL }\fi
\providecommand{\bibinfo}[2]{#2}
\providecommand{\eprint}[2][]{\url{#2}}

\bibitem[{\citenamefont{Walker et~al.}(2008)\citenamefont{Walker, McEwen, Le,
  Paolasini, and Fort}}]{hcwESRF07}
\bibinfo{author}{\bibfnamefont{H.~C.} \bibnamefont{Walker}},
  \bibinfo{author}{\bibfnamefont{K.~A.} \bibnamefont{McEwen}},
  \bibinfo{author}{\bibfnamefont{M.~D.} \bibnamefont{Le}},
  \bibinfo{author}{\bibfnamefont{L.}~\bibnamefont{Paolasini}},
  \bibnamefont{and} \bibinfo{author}{\bibfnamefont{D.}~\bibnamefont{Fort}},
  \bibinfo{journal}{J. Phys.: Condens. Matter} \textbf{\bibinfo{volume}{20}},
  \bibinfo{pages}{395221} (\bibinfo{year}{2008}).

\bibitem[{\citenamefont{Walker et~al.}(2007)\citenamefont{Walker, McEwen,
  Boulet, Colineau, Griveau, Rebizant, and Wastin}}]{hcw07}
\bibinfo{author}{\bibfnamefont{H.~C.} \bibnamefont{Walker}},
  \bibinfo{author}{\bibfnamefont{K.~A.} \bibnamefont{McEwen}},
  \bibinfo{author}{\bibfnamefont{P.}~\bibnamefont{Boulet}},
  \bibinfo{author}{\bibfnamefont{E.}~\bibnamefont{Colineau}},
  \bibinfo{author}{\bibfnamefont{J.-C.} \bibnamefont{Griveau}},
  \bibinfo{author}{\bibfnamefont{J.}~\bibnamefont{Rebizant}}, \bibnamefont{and}
  \bibinfo{author}{\bibfnamefont{F.}~\bibnamefont{Wastin}},
  \bibinfo{journal}{Phys. Rev. B} \textbf{\bibinfo{volume}{76}},
  \bibinfo{pages}{174437} (\bibinfo{year}{2007}).

\bibitem[{\citenamefont{Le et~al.}(2008{\natexlab{a}})\citenamefont{Le, Walker,
  McEwen, Gouder, Huber, and Wastin}}]{Gouder06}
\bibinfo{author}{\bibfnamefont{M.~D.} \bibnamefont{Le}},
  \bibinfo{author}{\bibfnamefont{H.~C.} \bibnamefont{Walker}},
  \bibinfo{author}{\bibfnamefont{K.~A.} \bibnamefont{McEwen}},
  \bibinfo{author}{\bibfnamefont{T.}~\bibnamefont{Gouder}},
  \bibinfo{author}{\bibfnamefont{F.}~\bibnamefont{Huber}}, \bibnamefont{and}
  \bibinfo{author}{\bibfnamefont{F.}~\bibnamefont{Wastin}},
  \bibinfo{journal}{J. Phys.: Condens. Matter} \textbf{\bibinfo{volume}{20}},
  \bibinfo{pages}{275220} (\bibinfo{year}{2008}{\natexlab{a}}).

\bibitem[{\citenamefont{Nellis et~al.}(1974)\citenamefont{Nellis, Harvey,
  Lander, Dunlap, Brodsky, Mueller, Reddy, and Davidson}}]{NHL+74}
\bibinfo{author}{\bibfnamefont{W.~J.} \bibnamefont{Nellis}},
  \bibinfo{author}{\bibfnamefont{A.~R.} \bibnamefont{Harvey}},
  \bibinfo{author}{\bibfnamefont{G.~H.} \bibnamefont{Lander}},
  \bibinfo{author}{\bibfnamefont{B.~D.} \bibnamefont{Dunlap}},
  \bibinfo{author}{\bibfnamefont{M.~B.} \bibnamefont{Brodsky}},
  \bibinfo{author}{\bibfnamefont{M.~H.} \bibnamefont{Mueller}},
  \bibinfo{author}{\bibfnamefont{J.~F.} \bibnamefont{Reddy}}, \bibnamefont{and}
  \bibinfo{author}{\bibfnamefont{G.~R.} \bibnamefont{Davidson}},
  \bibinfo{journal}{Phys. Rev. B} \textbf{\bibinfo{volume}{9}},
  \bibinfo{pages}{1041} (\bibinfo{year}{1974}).

\bibitem[{\citenamefont{Le et~al.}(2008{\natexlab{b}})\citenamefont{Le, McEwen,
  Wastin, Boulet, Colineau, Jardin, and Rebizant}}]{le07pupd3}
\bibinfo{author}{\bibfnamefont{M.~D.} \bibnamefont{Le}},
  \bibinfo{author}{\bibfnamefont{K.~A.} \bibnamefont{McEwen}},
  \bibinfo{author}{\bibfnamefont{F.}~\bibnamefont{Wastin}},
  \bibinfo{author}{\bibfnamefont{P.}~\bibnamefont{Boulet}},
  \bibinfo{author}{\bibfnamefont{E.}~\bibnamefont{Colineau}},
  \bibinfo{author}{\bibfnamefont{R.}~\bibnamefont{Jardin}}, \bibnamefont{and}
  \bibinfo{author}{\bibfnamefont{J.}~\bibnamefont{Rebizant}},
  \bibinfo{journal}{Physica B} \textbf{\bibinfo{volume}{403}},
  \bibinfo{pages}{1035} (\bibinfo{year}{2008}{\natexlab{b}}).

\bibitem[{\citenamefont{Bard}(1974)}]{bard}
\bibinfo{author}{\bibfnamefont{Y.}~\bibnamefont{Bard}},
  \emph{\bibinfo{title}{Nonlinear parameter estimation}}
  (\bibinfo{publisher}{Academic Press}, \bibinfo{year}{1974}), ISBN
  \bibinfo{isbn}{0120782502}.

\bibitem[{\citenamefont{Raphael and Lallement}(1968)}]{raphael_lallement}
\bibinfo{author}{\bibfnamefont{G.}~\bibnamefont{Raphael}} \bibnamefont{and}
  \bibinfo{author}{\bibfnamefont{R.}~\bibnamefont{Lallement}},
  \bibinfo{journal}{Sol. St. Comm.} \textbf{\bibinfo{volume}{6}},
  \bibinfo{pages}{383} (\bibinfo{year}{1968}).

\bibitem[{\citenamefont{McCart et~al.}(1981)\citenamefont{McCart, Lander, and
  Aldred}}]{mccart}
\bibinfo{author}{\bibfnamefont{B.}~\bibnamefont{McCart}},
  \bibinfo{author}{\bibfnamefont{G.~H.} \bibnamefont{Lander}},
  \bibnamefont{and} \bibinfo{author}{\bibfnamefont{A.~T.}
  \bibnamefont{Aldred}}, \bibinfo{journal}{J. Chem. Phys.}
  \textbf{\bibinfo{volume}{74}}, \bibinfo{pages}{5263} (\bibinfo{year}{1981}).

\bibitem[{\citenamefont{Carnall}(1992)}]{carnallAn}
\bibinfo{author}{\bibfnamefont{W.~T.} \bibnamefont{Carnall}},
  \bibinfo{journal}{J. Chem. Phys.} \textbf{\bibinfo{volume}{96}},
  \bibinfo{pages}{8713} (\bibinfo{year}{1992}).

\bibitem[{\citenamefont{Taylor et~al.}(1988)\citenamefont{Taylor, Osborn,
  McEwen, Stirling, Bowden, Williams, Balcar, and Lovesey}}]{taylorPrHET}
\bibinfo{author}{\bibfnamefont{A.~D.} \bibnamefont{Taylor}},
  \bibinfo{author}{\bibfnamefont{R.}~\bibnamefont{Osborn}},
  \bibinfo{author}{\bibfnamefont{K.~A.} \bibnamefont{McEwen}},
  \bibinfo{author}{\bibfnamefont{W.~G.} \bibnamefont{Stirling}},
  \bibinfo{author}{\bibfnamefont{Z.~A.} \bibnamefont{Bowden}},
  \bibinfo{author}{\bibfnamefont{W.~G.} \bibnamefont{Williams}},
  \bibinfo{author}{\bibfnamefont{E.}~\bibnamefont{Balcar}}, \bibnamefont{and}
  \bibinfo{author}{\bibfnamefont{S.~W.} \bibnamefont{Lovesey}},
  \bibinfo{journal}{Phys. Rev. Lett.} \textbf{\bibinfo{volume}{61}},
  \bibinfo{pages}{1309} (\bibinfo{year}{1988}).

\bibitem[{\citenamefont{Rotter}(2004)}]{rotter04}
\bibinfo{author}{\bibfnamefont{M.}~\bibnamefont{Rotter}}, \bibinfo{journal}{J.
  Magn. Mag. Mat.} \textbf{\bibinfo{volume}{272-276}}, \bibinfo{pages}{481}
  (\bibinfo{year}{2004}).

\bibitem[{\citenamefont{Newman and Ng}(2000)}]{NN00}
\bibinfo{author}{\bibfnamefont{D.~J.} \bibnamefont{Newman}} \bibnamefont{and}
  \bibinfo{author}{\bibfnamefont{B.~K.~C.} \bibnamefont{Ng}},
  \emph{\bibinfo{title}{Crystal Field Handbook}} (\bibinfo{publisher}{Cambridge
  University Press}, \bibinfo{year}{2000}).

\bibitem[{\citenamefont{Harvey et~al.}(1973)\citenamefont{Harvey, Brodsky, and
  Nellis}}]{harvey73}
\bibinfo{author}{\bibfnamefont{A.~R.} \bibnamefont{Harvey}},
  \bibinfo{author}{\bibfnamefont{M.~B.} \bibnamefont{Brodsky}},
  \bibnamefont{and} \bibinfo{author}{\bibfnamefont{W.~J.}
  \bibnamefont{Nellis}}, \bibinfo{journal}{Phys. Rev. B}
  \textbf{\bibinfo{volume}{7}}, \bibinfo{pages}{4137} (\bibinfo{year}{1973}).

\bibitem[{\citenamefont{Andersen and Smith}(1979)}]{nhandersen}
\bibinfo{author}{\bibfnamefont{N.} \bibnamefont{Hessel~Andersen}} \bibnamefont{and}
  \bibinfo{author}{\bibfnamefont{H.}~\bibnamefont{Smith}},
  \bibinfo{journal}{Phys. Rev. B} \textbf{\bibinfo{volume}{19}},
  \bibinfo{pages}{384} (\bibinfo{year}{1979}).

\bibitem[{\citenamefont{Gofryk et~al.}(2008)\citenamefont{Gofryk, Griveau,
  Colineau, and Rebizant}}]{gofryk_pupd5al2}
\bibinfo{author}{\bibfnamefont{K.}~\bibnamefont{Gofryk}},
  \bibinfo{author}{\bibfnamefont{J.-C.} \bibnamefont{Griveau}},
  \bibinfo{author}{\bibfnamefont{E.}~\bibnamefont{Colineau}}, \bibnamefont{and}
  \bibinfo{author}{\bibfnamefont{J.}~\bibnamefont{Rebizant}},
  \bibinfo{journal}{Phys. Rev. B} \textbf{\bibinfo{volume}{77}},
  \bibinfo{pages}{092405} (\bibinfo{year}{2008}).

\bibitem[{\citenamefont{Kondo}(1964)}]{kondores}
\bibinfo{author}{\bibfnamefont{J.}~\bibnamefont{Kondo}},
  \bibinfo{journal}{Progress of Theoretical Physics}
  \textbf{\bibinfo{volume}{32}}, \bibinfo{pages}{37} (\bibinfo{year}{1964}).

\bibitem[{\citenamefont{Rao and Wallace}(1970)}]{raowallace}
\bibinfo{author}{\bibfnamefont{V.~U.~S.} \bibnamefont{Rao}} \bibnamefont{and}
  \bibinfo{author}{\bibfnamefont{W.~E.} \bibnamefont{Wallace}},
  \bibinfo{journal}{Phys. Rev. B} \textbf{\bibinfo{volume}{2}},
  \bibinfo{pages}{4613} (\bibinfo{year}{1970}).

\bibitem[{\citenamefont{Mackintosh}(1963)}]{mackintosh63}
\bibinfo{author}{\bibfnamefont{A.~R.} \bibnamefont{Mackintosh}},
  \bibinfo{journal}{Phys. Lett.} \textbf{\bibinfo{volume}{4}},
  \bibinfo{pages}{140} (\bibinfo{year}{1963}).

\end{thebibliography}


\end{document}